\documentclass{eptcs}
\providecommand{\event}{DLS 2011} 

\usepackage[T1]{fontenc}
\usepackage[scaled=0.85]{beramono}
\usepackage{listings}
\usepackage{graphicx}

\title{Tutorial on Online Partial Evaluation}

\author{William R. Cook
       \institute{Department of Computer Science, University of Texas
             at Austin, USA}
       \and Ralf L\"ammel 
       \institute{Department of Computer Science, University of
            Koblenz-Landau, Germany}
         }
 
\def\titlerunning{Tutorial on Online Partial Evaluation}
\def\authorrunning{W. R. Cook \& R. L\"ammel}

\begin{document}
\maketitle

\begin{abstract}
  This paper is a short tutorial introduction to online partial
  evaluation. We show how to write a simple online partial evaluator
  for a simple, pure, first-order, functional programming language.
  In particular, we show that the partial evaluator can be derived as
  a variation on a compositionally defined interpreter. We demonstrate
  the use of the resulting partial evaluator for program optimization
  in the context of model-driven development.


\end{abstract}


\section{Introduction}

Partial evaluation is a powerful and general optimization technique~\cite{PartialEvalBookGomard}.
It has applications in model-driven development, domain-specific
language engineering~\cite{Hudak:1998:MDS:551789.853532}, 
and generic programming~\cite{Landauer:1999:GPP:874070.876061}.
The C++ template system can be understood as a form of partial
evaluation~\cite{Veldhuizen1998}. 
However, partial evaluation has not been very widely adopted.  It has
a reputation as an esoteric and complex topic. The challenge, set
forth by Futamura in the 1970s~\cite{Futamura+71}---to create an automatic
compiler-compiler by partial evaluation---has largely been abandoned; a
failure that may taint the entire effort.

These notes introduce partial evaluation by converting an interpreter
for a functional language into a simple partial evaluator for the
language.\footnote{A code distribution is available from the paper's
  website: \url{http://softlang.uni-koblenz.de/dsl11}} The conversion relies on a simple basic idea: 

\begin{quote}\itshape\centering
Allow undefined variables during evaluation.\\
Preserve syntactical phrases if they cannot be evaluated completely.
\end{quote}

\noindent
A regular interpreter maintains variable bindings in an
\textit{environment}.  Normally, the environment is required to
contain bindings for all variables mentioned in the program---an
undefined variable is a fatal error. For example, the evaluation of
the expression $x^n$ only succeeds if both $x$ and $n$ are bound to
values of types for which exponentiation is defined. Assuming bindings
$x=2$ and $n=3$, an interpreter evaluates $x^n$ to $8$, but it fails
if either $x$ or $n$ are not bound.

What would it mean to evaluate $x^n$ given a \textit{partial}
environment that only includes $n=3$?  The result cannot be a number,
because the value of $x$ is not known.  If we assume that the value of
$x$ will be known later, then the result can be a new program
$x^3$. This is called \textit{residual} code, because it is the code
that is left over after evaluating parts of the expression. Note that
the variable $n$ does not occur in the residual code. It has been
eliminated from the program.

Thus, one way to understand partial evaluation is this: enhance
regular evaluation to synthesize residual code when encountering
unbound variables in the partial environment. Partial evaluation
becomes more complicated when it deals with all the different kinds of
expressions that can occur in a program. Consider the following 
expression:

\begin{quote}
\lstinline{if x>y then (10+y)/x else y} 
\end{quote}

\noindent
What does it mean to partially evaluate the given expression in an
environment that specifies $y=0$ and with no binding for $x$? The
result is the residual code \lstinline{if x>0 then 10/x else 0}. In
this case, both branches of the conditional must be partially
evaluated, if $y$ is to be eliminated. This is not the normal
evaluation rule for conditionals.  As a result, partial evaluation may
diverge where regular evaluation would terminate---unless extra effort
is made. Also, great care must be taken in the presence of side
effects. Consider again the expression given earlier for a partial
environment such that $x=0$ and with no binding for $y$. Partial
evaluation is supposed to result in the residual code
\lstinline{if 0>y then error else y}. Note that the division-by-zero
error must not be raised during partial evaluation, but it must be
delayed so that it is executed at the right point in the residual
code---if it is ever exercised. Function calls and recursive
definitions also cause complications, which are discussed in the
tutorial.

The style of partial evaluator developed here is called an
\textit{online} partial evaluator, because it makes decisions about
specialization as it goes, based on whatever variables are in the
environment at a given point during evaluation~\cite{PartialEvalBookGomard}.  
An \textit{offline}
partial evaluator performs a static analysis of the program to decide
which variables will be considered known and which will be considered
unknown. It has been claimed 
that offline partial evaluation is simpler than online partial
evaluation~\cite{Thiemann95}. It can be more difficult to ensure 
termination in an online setting. In this tutorial, we avoid issues
of termination, so this complexity does not arise~\cite{Shali2011}.
Given this restriction, we believe that the online style is
simpler to describe, because the partial evaluator can be derived easily from an
existing interpreter.

\begin{figure}
\begin{lstlisting}[linerange=17-17]
type Prog = ([FDef], Expr)
\end{lstlisting} 
\begin{lstlisting}[linerange=21-21]
type FDef = (String,([String],Expr))
\end{lstlisting} 
\begin{lstlisting}[linerange=25-27]
data Val
 = IVal  { getInt :: Int }
 | BVal { getBool :: Bool }
\end{lstlisting} 
\begin{lstlisting}[linerange=35-40]
data Expr
 = Const { getVal :: Val }
 | Var String
 | Apply String [Expr]
 | Prim Op [Expr]
 | If Expr Expr Expr
\end{lstlisting} 
\begin{lstlisting}[linerange=62-62]
data Op = Equal | Add | Sub | Mul
\end{lstlisting} 
\caption{Syntax of a simple functional programming language}
\label{F:syntax}
\end{figure}

\section{A Simple Language and Evaluator}

Fig.~\ref{F:syntax} lists the abstract syntax for a simple, pure,
first-order, functional programming language. A program consists of a list of function definitions (see type
\lstinline{FDef}) and a main expression. A function definition is
tuple of the form $(f, ([a_1, ..., a_n], b))$ where $f$ is the function
name, $a_i$ are formal parameter names, and $b$ is the body expression
of the function. There are the following expression forms: a constant
(of type \lstinline{Val}), a variable, a function application, the
application of a primitive operator, and a conditional
expression. Evaluation of an expression is supposed to return a value,
but it may fail in reply to a dynamic type error; it may also diverge.
Functions are first-order: functions are neither passed as arguments,
nor returned as results.

As an example, consider the following exponentiation function written in Haskell:

\begin{lstlisting}
exp(x, n) = if n == 0 then 1 else x * exp(x, n - 1)
\end{lstlisting}

\noindent
This function can be encoded in our simple language as follows:

\begin{lstlisting}[linerange=157-163]
exp = (
 "exp", (["x","n"],
  If (Prim Equal [Var "n", Const (IVal 0)])
     (Const (IVal 1))  
     (Prim Mul 
       [Var "x",
        Apply "exp" [Var "x", Prim Sub [Var "n", Const (IVal 1)]]])))
\end{lstlisting} 

\noindent
The interpreter is a function \lstinline{eval} that takes a 
program (i.e., a list of function definitions and a main expression)
and returns a value (i.e., a Boolean or an int), or it fails, or it
diverges. For instance:

\begin{lstlisting}
> eval ([exp], Apply "exp" [Const (IVal 2), Const (IVal 3)])
8
\end{lstlisting}

\begin{figure}
\begin{lstlisting}[linerange=94-94]
type Env = [(String, Val)]
\end{lstlisting} 
\begin{lstlisting}[linerange=98-125]
eval :: Prog -> Val
eval (fdefs, main) = eval' main []

 where

  eval' :: Expr -> Env -> Val

  eval' (Const v) env = v
  
  eval' (Var s) env = 
    case lookup s env of
      Just v -> v
      Nothing -> error "undefined variable"
  
  eval' (Prim op es) env = 
    let rs = [ eval' e env | e <- es ] in
     prim op rs
  
  eval' (If e0 e1 e2) env =
    if getBool (eval' e0 env)
      then eval' e1 env
      else eval' e2 env
    
  eval' (Apply f es) env =
    eval' body env'
    where
      (ss, body) = fromJust (lookup f fdefs)
      env' = zip ss [ eval' e env | e <- es ]  
\end{lstlisting} 
\begin{lstlisting}[linerange=144-147]
prim Equal [IVal i1, IVal i2] = BVal (i1 == i2)
prim Add [IVal i1, IVal i2] = IVal (i1 + i2)
prim Sub [IVal i1, IVal i2] = IVal (i1 - i2)
prim Mul [IVal i1, IVal i2] = IVal (i1 * i2)
\end{lstlisting} 
\caption{An evaluator for Fig.~\ref{F:syntax}}
\label{F:Eval}
\end{figure}

\noindent
Fig.~\ref{F:Eval} gives the complete interpreter.  The main evaluation
function \lstinline{eval} binds the argument \lstinline{fdefs} to the list
of functions and it invokes the expression-level evaluation function
\lstinline{eval'} on the \lstinline{main} expression. Since
\lstinline{eval'} is defined in the scope of \lstinline{eval}, the
function list \lstinline{fdefs} does not need to be passed on
every recursive call. The helper function does take a
\textit{environment} \lstinline{env} that maps variables to values. The
environment is initialized to the empty list.

It must be emphasized that this interpreter is in no way special,
unusual, or targeted at partial evaluation. One gets this interpreter
naturally when using big-step style of operational semantics or direct
style of denotational semantics (i.e., compositional, functional
semantics) as the guiding principle.

\section{A Naive Partial Evaluator}

Let us start the conversion of the interpreter into a partial
evaluator. The partial evaluator should produce the same result as the
interpreter when given a complete environment, but it should compute a
residual expression when given a partial environment. Let us think of
the residual expression as the ``rest of the program'' that needs to
be evaluated once the remaining variables are bound to values. For
instance:

\begin{lstlisting}
> peval ([exp], Apply "exp" [Var "x", Const (IVal 3)])

Prim Mul [Var "x", 
              Prim Mul [Var "x",
                            Prim Mul [Var "x", Const (IVal 1)]]]
\end{lstlisting}

\noindent
Thus, the function \lstinline{exp} is applied to a specific exponent
$3$, but the base remains a variable $x$, and the result of partial
evaluation represents the expression to compute $x^3 = x*x*x*1$. The
partial evaluator derived that expression by applying (unfolding) the
recursive definition of \lstinline{exp} for exponents 3, 2, 1, and 0.

We develop a partial evaluator in two steps. The intermediate version
is very simple, but also limited. In fact, it is essentially a
systematic inlining transformation. We start from the insight that
the type of the top-level function must be changed so that expressions
are returned instead of values. Please note that values are trivially
embedded into expressions through the constant form of
expressions. Thus:

\begin{lstlisting}
-- Interpreter
eval :: Prog -> Val

-- Partial evaluator
peval :: Prog -> Expr
\end{lstlisting}

\noindent
Also, our initial, naive development uses the idea that function
applications should always be inlined, and hence, we need to be able
to apply functions to results of partial evaluation, i.e., to
expressions rather than values. Accordingly, the type of environments
changes as follows:

\begin{lstlisting}
-- Interpreter
type Env = [(String, Val)]

-- Naive, inlining-oriented partial evaluator
type Env = [(String, Expr)]
\end{lstlisting}

\begin{figure}
\begin{lstlisting}[linerange=23-23]
type Env = [(String, Expr)]
\end{lstlisting} 
\begin{lstlisting}[linerange=27-61]
peval :: Prog -> Expr
peval (fdefs, main) = peval' main []

 where

  peval' :: Expr -> Env -> Expr

  peval' (Const v) env = Const v

  peval' (Var s) env =
    case lookup s env of
      Just e  -> e
      Nothing -> Var s -- return code for variable

  peval' (Prim op es) env =
   let rs = [ peval' e env | e <- es ] in
    if all isVal rs
     then Const (prim op (map getVal rs))
     else Prim op rs -- return code for primitive

  peval' (If e0 e1 e2) env =
   let r0 = peval' e0 env in
    if isVal r0 
     then if getBool (getVal r0) 
           then peval' e1 env
           else peval' e2 env
     else (If r0 
                (peval' e1 env) 
                (peval' e2 env)) -- return code for if

  peval' (Apply f es) env =
   peval' body env'
   where
    (ss, body) = fromJust (lookup f fdefs)
    env' = zip ss [ peval' e env | e <- es ]  
\end{lstlisting} 
\caption{The naive partial evaluator}
\label{F:SimplePEval}
\end{figure}

\noindent
Fig.~\ref{F:SimplePEval} gives the complete, naive partial
evaluator. Function \lstinline{peval'} generalizes \lstinline{eval'} as follows:

\begin{itemize}

\item A \emph{constant} is partially evaluated to itself (just like in
  the interpreter).

\item A \emph{variable} is partially evaluated to the value (constant)
  according to the variable's binding in the environment (just like in
  the interpreter), if there is a binding. Otherwise, the variable is
  partially evaluated to itself. (The interpreter failed in this case.)

\item A \emph{primitive} operation is applied to the (partially)
  evaluated arguments (just like in the interpreter), if these are all
  values (constants). Otherwise, a primitive operation form of
  expression is reconstructed from the partially evaluated arguments.

\item A \emph{conditional} can be eliminated such that one of the two
  branches is chosen for recursive (partial) evaluation (just like in
  the interpreter), if the condition is (partially) evaluated to a
  value (constant). Otherwise, both branches are partially evaluated,
  and the conditional is reconstructed.

\item A \emph{function application} is partially evaluated just like
  in the interpreter---modulo the changed environment type. (Alpha
  renaming should be applied to avoid any name confusion but this
  is omitted here for brevity.)

\end{itemize}

The treatment of conditionals and function applications is naive. For
example, partial evaluation of a function application may diverge when
compared to regular evaluation. Thus:

\begin{lstlisting}
> peval ([exp], Apply "exp" [Const (IVal 2), Var "n"])

-- Result shown in regular Haskell notation for clarity
if n == 0
  then 1
  else 2 * (if n-1 == 0
                  then 1 
                  else 2 * (if (n-1)-1 == 0 ...))
\end{lstlisting}

\noindent
In this example, the function \lstinline{exp} is applied to a specific
base $2$, but the exponent remains a variable $n$.  Inlining diverges
because the recursive case of \lstinline{exp} is continuously exercised
for different argument expressions for the exponent.

\section{Proper Program Specialization}

Proper treatment of recursive functions requires from us to
synthesize \emph{residual programs} instead of just residual
expressions based on naive inlining. Hence, our more advanced partial
evaluator uses the following type:

\begin{lstlisting}
peval :: Prog -> Prog
\end{lstlisting}

\noindent
Also, we return to the original definition of \lstinline{Env}, which
binds variables to values rather than expressions. The idea is here
that the incoming function definitions and the main expression are
specialized such that the resulting main expression only refers to
specialized function definitions. The same original function
definition may be specialized several times depending on the
encountered, statically known argument values. For instance:

\begin{lstlisting}
> peval ([exp], Apply "exp" [Var "x", Const (IVal 3)])

([ ("exp'a",(["x"],
       Prim Mul [Var "x", Apply "exp'b" [Var "x"]])),
   ("exp'b",(["x"],
       Prim Mul [Var "x", Apply "exp'c" [Var "x"]])),
   ("exp'c",(["x"],
       Prim Mul [Var "x", Apply "exp'd" [Var "x"]])),
   ("exp'd",(["x"],
       Const (IVal 1))), 
 ],
 Apply "exp'a" [Var "x"]
)
\end{lstlisting}

\noindent
The names of the specialized functions are fabricated from the
original name by some qualification scheme. For clarity, the list of
function definitions are also shown in plain Haskell:

\begin{lstlisting}
exp'a x = x * exp'b x
exp'b x = x * exp'c x
exp'c x = x * exp'd x
exp'd x = 1
\end{lstlisting}

\noindent
Thus, specialized function definitions were inferred for all the
inductively encountered values 3, 2, 1, and 0 for the exponent. Thus,
subject to a simple inlining optimization, which is not shown here for
brevity, we obtain the familiar expression for \lstinline{x} to the
power 3.

Let us apply the more advanced, partial evaluator to the diverging
example that we faced at the end of the previous section. Function
specialization carefully tracks argument lists for which
specialization is under way or has been completed. This solves
the termination problem.

\begin{lstlisting}
> peval ([exp], Apply "exp" [Const (IVal 2), Var "n"])

(["exp'", (["n"],
  If (Prim Equal [Var "n", Const (IVal 0)])
     (Const (IVal 1))  
     (Prim Mul 
        [Const (IVal 2),
         Apply "exp'" [Prim Sub [Var "n", Const (IVal 1)]]]))],
 Apply "exp'" [Var "n"]
)
\end{lstlisting}

\noindent
Thus, the original definition of \lstinline{exp} was specialized
such that the argument position for the statically known base is
eliminated. Please note that the specialized function is recursive.

The partial evaluator needs to aggregate specialized functions along
with recursion into expressions. To this end, we use the state monad
in the type of the expression-level function \lstinline{peval'}. Thus:

\begin{lstlisting}
-- Naive inlining transformation
peval' :: Expr -> Env -> Expr

-- Proper program specialization
peval' :: Expr -> Env -> State [FDef] Expr
\end{lstlisting}

\begin{figure}
\begin{lstlisting}[linerange=94-94]
type Env = [(String, Val)]
\end{lstlisting} 
\begin{lstlisting}[linerange=23-28]
peval :: Prog -> Prog
peval (fdefs, main) = swap (runState (peval' main []) [])

 where

  peval' :: Expr -> Env -> State [FDef] Expr 
\end{lstlisting} 
\begin{lstlisting}[linerange=32-54]
  peval' (Const v) env = return (Const v)

  peval' (Var s) env = 
    case lookup s env of
      Just v  -> return (Const v)
      Nothing -> return (Var s)

  peval' (Prim op es) env = do
    rs <- mapM (flip peval' env) es
    if all isVal rs
     then return (Const (prim op (map getVal rs)))
     else return (Prim op rs)

  peval' (If e0 e1 e2) env = do 
    r0 <- peval' e0 env
    if isVal r0 then
      if getBool (getVal r0) 
       then peval' e1 env
       else peval' e2 env
     else do 
      r1 <- peval' e1 env
      r2 <- peval' e2 env
      return (If r0 r1 r2)
\end{lstlisting} 
\caption{Monadic partial evaluation}
\label{F:State}
\bigskip
\end{figure}

\noindent
The cases for all constructs but function application can be adopted
from the simpler partial evaluator---except that we need to convert to
monadic style, which is a simple, systematic program transformation in
itself~\cite{Laemmel99,ErwigR04}. See Fig.~\ref{F:State} for the
result. That is, recursive calls to \lstinline{peval'} are not used
directly in reconstructing terms, but their results are bound in the
state monad.

\begin{figure}
\begin{lstlisting}[linerange=66-104]
  peval' (Apply s es) env = do

    -- Look up function.
    let (ss, body) = fromJust (lookup s fdefs)

    -- Partially evaluate arguments.
    rs <- mapM (flip peval' env) es

    -- Determine static and dynamic arguments.
    let z = zip ss rs
    let sas = [ (s, getVal r) | (s, r) <- z, isVal r ]
    let das = [ (s,v) | (s,v) <- z, not (isVal v) ]

    if null das then 

      -- Inline completely static applications.
      peval' body sas 
  
      -- Otherwise these applications make their contexts dynamic.
     else do

        -- Fabricate name from static variables.
        let s' = s ++ show (hashString (show sas))

        -- Specialize each "name" just once.
        fdefs <- get
        when (isNothing (lookup s' fdefs)) (do

           -- Create placeholder for memoization.
           put (fdefs ++ [(s', undefined)]) 

           -- Partially evaluate function body.
           e' <- peval' body sas

           -- Replace placeholder by actual definition.
           modify (update (const (map fst das, e')) s'))

        -- Return application of specialized function.
        return (Apply s' (map snd das))
\end{lstlisting} 
\caption{Partial evaluation of function application}
\label{F:Apply}
\end{figure}

It remains to define the case for partial evaluation of function
application. We provide an informal specification for this case;
please refer to Fig.~\ref{F:Apply} for the actual implementation:

\begin{enumerate}

\item The applied function is looked up and the arguments
are evaluated---just like in the interpreter. 

\item The partially evaluated arguments are partitioned into static
  and dynamic ones.  Static arguments are values (constants); dynamic
  arguments leverage other expression forms.

\item The ``identity'' (the name) of the specialized function derives
  from the applied function and the static arguments. Here, we assume
  that values can be compared for equality. This is essential for
  remembering (say, memoizing) previous function specializations.

\item The body of the specialized function is obtained by partially
  evaluating the original body in the variable environment of the
  static variables. The argument list of the specialized function only
  includes variables for the dynamic positions.

\item The specialized function is ultimately applied to the dynamic
  arguments. The expression for that application serves as the result 
  of partial evaluation.
 
\end{enumerate}

Close inspection of Fig.~\ref{F:Apply} reveals additional details.
For the special case of an application with static arguments only (see
\lstinline{if null das then ...}), we switch to the behavior of the
interpreter by (partially) evaluating the body of the applied
function. Also, in order to deal with recursion, it is important that
the specialized function is already added to the state before its body
is obtained. To this end, an undefined body is initially registered as
a placeholder to be updated later.

\section{Applications of Partial Evaluation}

The exponentiation function is used frequently in discussing partial
evaluation. Such simple examples are useful, because they illustrate
the basic ideas with as little complication as possible. However,
there is not much benefit in partially evaluating simple numeric
functions, like exponentiation. Thus the potential benefits of the
approach are not highlighted.

One promising application area for partial evaluation is in model-driven
development. A \textit{model} is a description of some desired behavior.
While it is common practice to \textit{translate} models into code that
generates the desired behavior, we consider another approach.
A \textit{model interpreter} takes the model as an input and performs
the behaviors specified in the model. Using partial evaluation, we can
specialize a model interpreter to create a compiled 
version of a model~\cite{Futamura+71}.
To illustrate this technique, we show how to partially evaluate
a state machine interpreter written in our simple language.
In what follows, we use Haskell code to represent the interpreter,
although it can be easily translated into our simple language for
interpretation and partial evaluation.

\begin{figure}[t!]
\begin{lstlisting}[linerange=16-19]
type State = Int
type Label = Char
type Transitions = [(State, [(Label, State)])]
type Accept = [State]
\end{lstlisting} 
\begin{center}
\includegraphics[scale=0.25]{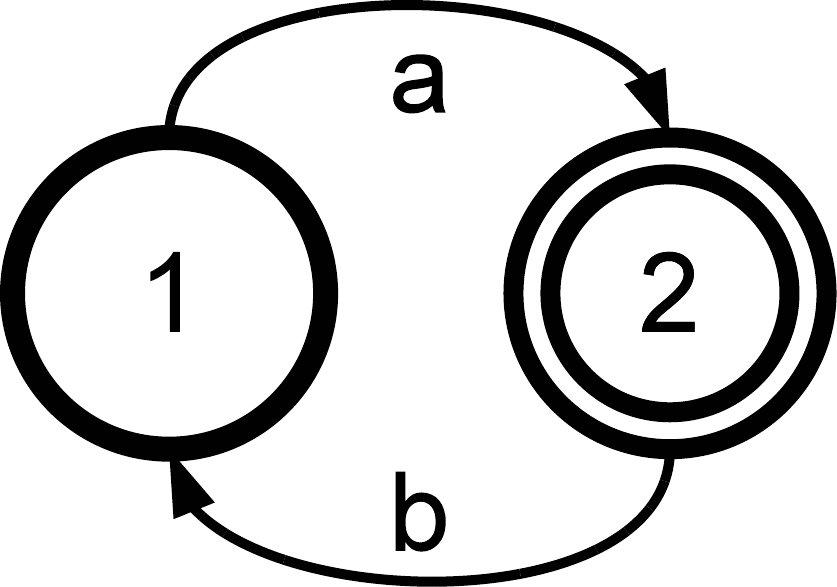}
\end{center}
\begin{lstlisting}[linerange=37-39]
transitions = [(1, [('a', 2)]), 
                    (2, [('b', 1)])]
accept = [2]
\end{lstlisting} 
\caption{Types for state machines and example\label{dfaex}}
\bigskip
\end{figure}

Fig.~\ref{dfaex} defines a type for representing state machines, and
gives a simple state machine with two states. (We assume deterministic
finite automa here.) A state machine is encoded using two Haskell data
types for the accept states and the transition table. The double
circle on state 2 indicates that it is the accept state.

\begin{figure}[t!]
\begin{lstlisting}[linerange=25-30]
run :: State -> Accept -> Transitions -> [Label] -> Bool
run current accept transitions [] = current `elem` accept
run current accept transitions (l:ls) =
  case lookup l (fromJust (lookup current transitions)) of
    Nothing -> False
    Just next -> run next accept transitions ls
\end{lstlisting} 
\caption{Haskell interpreter for state machines\label{dfaint}}
\bigskip
\end{figure}

\begin{figure}[t!]
\begin{lstlisting}[linerange=4-14]
run1 ls = if null ls
                then False
                else if head ls == 'a'
                          then run2 (tail ls)
                          else False
       
run2 ls = if null ls
                then True
                else if head ls == 'b'
                          then run1 (tail ls)
                          else False
\end{lstlisting} 
\caption{Desired output from partial evaluation\label{dfaout}}
\bigskip
\end{figure}

Fig.~\ref{dfaint} gives a Haskell version of a simple interpreter for
state machines. The interpreter takes the current state, the accept
states, the transition table, and a sequence of input labels. It
returns true if the machine consumes all input and ends in an accept
state.

To translate the interpreter into our simple language, additional data
structures and primitive operators must be added to the language, to
represent pairs and lists as well as maybes. These changes are left as
an exercise for the reader. The \lstinline{lookup} and
\lstinline{elem} function must also be written.

The desired result from partially evaluating the state machine
interpreter on the state machine in Fig.~\ref{dfaex} is given in
Fig.~\ref{dfaout}. The accept states and the transition table are no
longer present as data structures---thereby promising an aggressive
optimization. However, when our partial evaluator is applied to the
program in Fig.~\ref{dfaex} (appropriately encoded in our simple
language), specialization fails to eliminate the data structures for
accept states and transition table. The problem is this expression:

\begin{quote}
\lstinline{lookup current transitions}
\end{quote}

\noindent
In this call, the \lstinline{current} variable is \textit{dynamic}
during partial evaluation, although \lstinline{transitions} is
static. Given a dynamic input, the result of
\lstinline{lookup current transitions} is always dynamic (that is,
unknown) even though the result is a list of transitions which is
static information in the program. Thus, partial evaluation stops
prematurely.

This is a well-known problem in partial evaluation: when static and
dynamic information are tangled together, the static information is lost
and partial evaluation stops. In such cases, it is necessary to rewrite 
the program so that it works better with the partial evaluator. Such
rewrites are known as ``binding time improvements''~\cite{PartialEvalBookGomard}.
There is a huge body of work on this problem, but our example
provides a simple illustration. The improvement
pattern we need can be described with the following pseudo code.
Our current program works by looking up the dynamic key in a static table, then
processing the result:

\begin{lstlisting}
let v = lookup table key in
  process v
\end{lstlisting}

Instead, we must iterate over the static table, and compare its keys
to the dynamic key, and then process the value, if it matches:

\begin{lstlisting}
for (k,v) in table do
  if k == key then
    process v
\end{lstlisting}

A partial evaluator can unroll this loop such that \lstinline{v} is
static in the call \lstinline{process v} and partial evaluation will
continue. In our functional language, we do not use any loop
construct, but we use recursion instead. Fig.~\ref{dfafinal} gives a
rewritten version of the state machine interpreter with the binding
time improvement. This version is also more directly translatable to
our simple language, because it does not use pattern matching.

\begin{figure}[t!]
\begin{lstlisting}[linerange=17-28]
run :: State -> Accept -> Transitions -> [Label] -> Bool
run s accept transitions ls =
  if null ls
  then s `elem` accept
  else runNext (lookup s transitions) accept transitions ls

runNext s accept transitions ls =
  if null s 
  then False
  else if (head ls) == fst (head s)
          then run (snd (head s)) accept transitions (tail ls)
          else runNext (tail s) accept transitions ls
\end{lstlisting} 
\caption{State machine interpreter with binding time improvements}
\label{dfafinal}
\bigskip
\end{figure}

\section{Further Reading}

These tutorial notes are only a quick introduction to the idea of partial 
evaluation. There are many sources for further reading. 
We recommend Jones, Gomard and Sestoft's book on partial
evaluation~\cite{PartialEvalBookGomard} and 
John Hatcliff's detailed course material~\cite{Hatcliff_foundationsfor}
as a good place to start. Both are available online.

\section{Conclusion}\label{conclusion}

This introduction merely hints at a number of important questions
and issues. For example, our partial evaluator does not always terminate.
Ensuring termination is a complex problem, but it is also possible to 
leave it to the programmer to avoid termination problems. 
Controlling partial evaluation is an important area of research~\cite{LeMeur:2002:TBG:503032.503033}.
A partial evaluator can fail to specialize 
desired parts of a program (as illustrated in the last section), or 
it can generate huge amounts of code without any practical benefit. 
Is it possible to automatically refactor programs so that
programs will work better with partial evaluation?
Dealing with imperative effects (e.g., mutable variables and input/output)
is also a significant problem. We hope that this tutorial will 
encourage more researchers to investigate and apply partial evaluation 
in their work.

\bibliographystyle{eptcs}
\bibliography{cook,laemmel}

\begin{thebibliography}{10}
\providecommand{\bibitemdeclare}[2]{}
\providecommand{\urlprefix}{Available at }
\providecommand{\url}[1]{\texttt{#1}}
\providecommand{\href}[2]{\texttt{#2}}
\providecommand{\urlalt}[2]{\href{#1}{#2}}
\providecommand{\doi}[1]{doi:\urlalt{http://dx.doi.org/#1}{#1}}
\providecommand{\bibinfo}[2]{#2}

\bibitemdeclare{article}{ErwigR04}
\bibitem{ErwigR04}
\bibinfo{author}{Martin Erwig} \& \bibinfo{author}{Deling Ren}
  (\bibinfo{year}{2004}): \emph{\bibinfo{title}{Monadification of functional
  programs}}.
\newblock {\sl \bibinfo{journal}{Sci. Comput. Program.}} \bibinfo{volume}{52},
  pp. \bibinfo{pages}{101--129}, \doi{10.1016/j.scico.2004.03.004}.

\bibitemdeclare{article}{Futamura+71}
\bibitem{Futamura+71}
\bibinfo{author}{Yoshihiko Futamura} (\bibinfo{year}{1999}):
  \emph{\bibinfo{title}{Partial Evaluation of Computation Process --- An
  Approach to a Compiler-Compiler}}.
\newblock {\sl \bibinfo{journal}{Higher Order Symbol. Comput.}}
  \bibinfo{volume}{12}, pp. \bibinfo{pages}{381--391},
  \doi{10.1023/A:1010095604496}.

\bibitemdeclare{misc}{Hatcliff_foundationsfor}
\bibitem{Hatcliff_foundationsfor}
\bibinfo{author}{John Hatcliff} (\bibinfo{year}{1999}):
  \emph{\bibinfo{title}{Foundations of Partial Evaluation and Program
  Specialization}}.
\newblock \urlprefix\url{http://people.cis.ksu.edu/~hatcliff/FPEPS/}.

\bibitemdeclare{inproceedings}{Hudak:1998:MDS:551789.853532}
\bibitem{Hudak:1998:MDS:551789.853532}
\bibinfo{author}{P.~Hudak} (\bibinfo{year}{1998}):
  \emph{\bibinfo{title}{Modular Domain Specific Languages and Tools}}.
\newblock In: {\sl \bibinfo{booktitle}{Proceedings of the 5th International
  Conference on Software Reuse}}, \bibinfo{series}{ICSR '98},
  \bibinfo{publisher}{IEEE Computer Society}, \bibinfo{address}{Washington, DC,
  USA}, pp. \bibinfo{pages}{134--}.

\bibitemdeclare{book}{PartialEvalBookGomard}
\bibitem{PartialEvalBookGomard}
\bibinfo{author}{Neil~D. Jones}, \bibinfo{author}{Carsten~K. Gomard} \&
  \bibinfo{author}{Peter Sestoft} (\bibinfo{year}{1993}):
  \emph{\bibinfo{title}{Partial evaluation and automatic program generation}}.
\newblock \bibinfo{publisher}{Prentice-Hall, Inc.}, \bibinfo{address}{Upper
  Saddle River, NJ, USA}.

\bibitemdeclare{inproceedings}{Laemmel99}
\bibitem{Laemmel99}
\bibinfo{author}{Ralf L{\"a}mmel} (\bibinfo{year}{1999}):
  \emph{\bibinfo{title}{{Reuse by Program Transformation}}}.
\newblock In: {\sl \bibinfo{booktitle}{{Selected papers from the 1st Scottish
  Functional Programming Workshop (SFP 1999)}}}, {\sl \bibinfo{series}{Trends
  in Functional Programming}}~\bibinfo{volume}{1},
  \bibinfo{publisher}{Intellect}, pp. \bibinfo{pages}{144--153}.

\bibitemdeclare{inproceedings}{Landauer:1999:GPP:874070.876061}
\bibitem{Landauer:1999:GPP:874070.876061}
\bibinfo{author}{C.~Landauer} \& \bibinfo{author}{K.L. Bellman}
  (\bibinfo{year}{1999}): \emph{\bibinfo{title}{Generic programming, partial
  evaluation, and a new programming paradigm}}.
\newblock In: {\sl \bibinfo{booktitle}{System Sciences, 1999. HICSS-32.
  Proceedings of the 32nd Annual Hawaii International Conference on}},
  \bibinfo{volume}{Track3}, p. \bibinfo{pages}{10 pp.},
  \doi{10.1109/HICSS.1999.772896}.

\bibitemdeclare{inproceedings}{LeMeur:2002:TBG:503032.503033}
\bibitem{LeMeur:2002:TBG:503032.503033}
\bibinfo{author}{Anne-Fran\c{c}oise Le~Meur}, \bibinfo{author}{Julia~L. Lawall}
  \& \bibinfo{author}{Charles Consel} (\bibinfo{year}{2002}):
  \emph{\bibinfo{title}{Towards bridging the gap between programming languages
  and partial evaluation}}.
\newblock In: {\sl \bibinfo{booktitle}{Proceedings of the 2002 ACM SIGPLAN
  workshop on Partial evaluation and semantics-based program manipulation}},
  \bibinfo{series}{PEPM '02}, \bibinfo{publisher}{ACM}, pp.
  \bibinfo{pages}{9--18}, \doi{10.1145/503032.503033}.

\bibitemdeclare{inproceedings}{Shali2011}
\bibitem{Shali2011}
\bibinfo{author}{Amin Shali} \& \bibinfo{author}{William~R. Cook}
  (\bibinfo{year}{2011}): \emph{\bibinfo{title}{Hybrid Partial Evaluation}}.
\newblock In: {\sl \bibinfo{booktitle}{Proc. of ACM Conf. on Object-Oriented
  Programming, Systems, Languages and Applications}}.
\newblock \bibinfo{note}{(to appear)}.

\bibitemdeclare{inproceedings}{Thiemann95}
\bibitem{Thiemann95}
\bibinfo{author}{Peter Thiemann} \& \bibinfo{author}{Robert Gl{\"u}ck}
  (\bibinfo{year}{1995}): \emph{\bibinfo{title}{The Generation of a
  Higher-Order Online Partial Evaluator}}.
\newblock In: {\sl \bibinfo{booktitle}{Fuji Workshop on Functional and Logic
  Programming}}, pp. \bibinfo{pages}{239--253}.

\bibitemdeclare{inproceedings}{Veldhuizen1998}
\bibitem{Veldhuizen1998}
\bibinfo{author}{Todd~L. Veldhuizen} (\bibinfo{year}{1999}):
  \emph{\bibinfo{title}{C++ Templates as Partial Evaluation}}.
\newblock In: {\sl \bibinfo{booktitle}{Partial Evaluation and Semantic-Based
  Program Manipulation}}, pp. \bibinfo{pages}{13--18}.

\end{thebibliography}

\end{document}